\documentstyle[12pt]{article}

\def\as{\alpha_{\rm S}}

\def\n{\!\!}

\def\a{\alpha}

\def\D{\Delta}

\def\n{\nu}

\def\pa{\partial}

\def\ra{\rightarrow}

\def\ti{\tilde}

\def\({\left(}
\def\){\right)}

\def\citenum#1{{\def\@cite##1##2{##1}\cite{#1}}}
\def\citea#1{\@cite{#1}{}}

\def\beq{\begin{equation}}
\def\eeq{\end{equation}}
\def\bea{\begin{eqnarray}}
\def\eea{\end{eqnarray}}
\def\eq#1{{eq.~(\ref{#1})}}
\def\Eq#1{{Eq.~(\ref{#1})}}

\def\bbbz{{\mathchoice {\hbox{$\sf\textstyle Z\kern-0.4em Z$}}
{\hbox{$\sf\textstyle Z\kern-0.4em Z$}}
{\hbox{$\sf\scriptstyle Z\kern-0.3em Z$}}
{\hbox{$\sf\scriptscriptstyle Z\kern-0.2em Z$}}}}
\relax

\begin{document}
\begin{flushright}
CERN-TH/95-61\\
TAUP-2226-95\\
CBPF NF-012/95\\
March 1995
\end{flushright}
\vskip 0.7 true cm
\begin{center}
\large {\bf A NEW EVOLUTION EQUATION}

 \end{center}
\vskip 12pt

\begin{center}
\begin{sc}
 Eric Laenen
\end{sc}
\footnote{{\tt laenen@surya11.cern.ch}}
 \par \vskip 0.2 true cm
 \noindent
 {\sl CERN Theory Division\\
   CH-1211, Geneva 23\\
   Switzerland}
\vskip 0.6cm

\begin{sc}
Eugene Levin
\end{sc}
\footnote{{\tt levin@lafex.cbpf.br; levin@ccsg.tau.ac.il}}
\footnote{On leave from
{\sl
Theory Department,
St. Petersburg Nuclear Physics Institute,
188350, St. Petersburg, Gatchina, Russia}}
\par \vskip 0.2 true cm
{\sl Mortimer and Raymond Sackler Institute of Advanced Studies\\
School of Physics and Astronomy, Tel Aviv University\\
Ramat Aviv, 69978, Israel}
 \\ and\\
{\sl LAFEX, Centro Brasileiro de Pesquisas F\'\i sicas / CNPq\\
Rua Dr. Xavier Sigaud 150, 22290 - 180, Rio de Janiero, RJ, Brasil}

\noindent
\end{center}
\par \vskip .1in

\begin{abstract}
\par \vskip .2in \noindent
We propose a new evolution equation for the gluon density relevant
for the region of small $x_B$. It generalizes the
GLR equation and allows deeper penetration in dense parton
systems than the GLR equation does. This generalization consists
of taking shadowing effects more comprehensively into account
by including multigluon correlations, and allowing for an
arbitrary initial gluon distribution in a hadron. We solve the
new equation for fixed $\alpha_s$. We find that the effects of
multigluon correlations on the deep-inelastic structure function
are small.
\end{abstract}



\pagebreak
\section{Introduction}
Our objective in this paper is to derive a new evolution equation
describing the behavior of large partonic densities. The fact that
parton densities increase as the Bjorken scaling variable
$x_B$ decreases follows directly from linear evolutions equations
such as the GLAP \cite{GLAP} or BFKL \cite{BFKL} equation,
and has been experimentally observed by both HERA collaborations
\cite{HERA}.
They observe a powerlike behavior of the deep inelastic structure
function
\beq \label{ONE}
F_2 (x_B,Q^2) \,\,\sim \,\,x^{- \omega_0}_B,
\eeq
with $\omega_0 \simeq  0.3 -  0.5$. This
behavior is predicted by the BFKL equation and is not inconsistent with
the GLAP equation. (Such powerlike growth at
small $x$ can be imitated by a solution of the GLAP
equation with a distribution that is constant in
$x$ at a low initial scale $Q_0$.)

{}From a physical point of view such behavior is inconsistent
with unitarity. This fact necessitates a change in the evolution
equation in the region of small $x_B$. The first attempt, more than
ten years old, to write down a new equation led to the nonlinear
GLR equation \cite{GLR}. See \cite{REVIEW} for a more extensive
review.

We will explain in this paper that the GLR equation only includes two-gluon
correlations in the parton cascade. Multigluon
correlations should be essential to solving the small $x_B$ problem,
at least theoretically \cite{TWOTW,HIGHTW}. It is the aim of
this paper to take such correlations into account.

In section 2 we derive and explain the limitations of the GLR
equation. We consider multigluon correlations in section 3 by
employing the relation between these correlations and high twist
contributions to the deep-inelastic structure function.
In section 4 we suggest the evolution equation which takes
these correlations into account. Some particular solutions
and a comparison with the GLR equation are discussed in section 5, while
we discuss the general solution of the new evolution equation in section 6.
We summarize and conclude in section 7.

\section{The GLR Equation}

In the region of small $x_B$ and large $Q^2$ we face a system of
partons at mutually small distances (thus the QCD coupling $\alpha_s$
is still small), but dense enough that the usual perturbative
QCD (pQCD) methods cannot be applied. The physics that governs this
region is nonperturbative, but of a different nature than the
one
associated with large distances. The latter corresponds to the
confinement region, and is usually analyzed using lattice field
theory or QCD sum rules. In contrast, we encounter here a situation
where new methods must be devised to analyze such a dense relativistic
system of partons in a non-equilibrium state. We need, in fact, new
quantum statistical methods to describe the behavior of such a
system and to chart this unknown region. We are unfortunately
only at the beginning of this exploration.

On the upside, we can approach this kinematical region in theory
from the pQCD region, and assume that in a transition region between
pQCD and high density QCD (hdQCD) we can study such a dense system
in some detail. To illustrate what new physics one might anticipate
in this transition region let us look at deep-inelastic scattering
and compare with the pQCD results for this process. We can
expect the following phenomena in the transition region:

(i) As $x_B$ decreases the total cross section
$\sigma (\gamma^* N)$ grows and, near the border with the hdQCD domain,
becomes even comparable with the geometrical size of the nucleon
$\sigma(\gamma^* N) \rightarrow \alpha_{em}
\pi R^2_N$. Here the cross section should be a smooth function of
$\ln Q^2$.

(ii) Although the parton language can be used to discuss the main properties
of the process, interactions between partons become
important, especially the annihilation process. This interaction
induces screening (a.k.a. shadowing) corrections.

(iii) In this particular kinematical region such screening corrections are
fortunately under theoretical control. They modify however the
pQCD linear evolution equation.
The correct evolution equation now becomes nonlinear.

First some nomenclature.
In this paper we will only deal with the gluon density
in the nucleon, $x_B G(x_B,Q^2)$, which can be measured
fairly directly at small $x_B$ in e.g experiments on diffraction
dissociation or heavy quark production in deeply inelastic
processes or even directly from the deep-inelastic
structure function $F_2(x_B,Q^2)$ \cite{EKL}.
We will thus also refer to it as the gluon structure
function.

We will now give a simple derivation of this nonlinear equation,
based on physical concepts.
As we stated, the main new processes that we must consider at high
density are parton-parton interactions. To incorporate these in our
physical descriptions we must identify a new small parameter that
controls the accuracy of calculations involving these
interactions. Such a parameter is
\beq \label{W}
 W\,\,=\,\,\frac{\alpha_s}{Q^2} \cdot \rho\,\,,
\eeq
where $\rho$ is the gluon density in the transverse plane
\beq
\rho\,\,=\,\,\frac{x_B G (x_B,Q^2)}{\pi R_N^2}\,\,.
\eeq
Here $R_N$ characterizes the area of a hadron which the gluons
populate. It is the correlation length of gluons inside
a hadron. Naturally, this radius must be smaller than the radius
of a hadron (proton). Since this paper is mostly devoted to the
discussion of purely theoretical questions we will not specify further the
value of $R_N$. However it should be stressed that $R_N$ is
nonperturbative in nature: all physics that occurs at distance
scales larger than $R_N$ is nonperturbative.

The first factor in
(\ref{W}) is the cross section for gluon absorption by a parton
in the hadron. Hence $W$ has the simple physical meaning of
being the probability of parton (mainly gluon) recombination in the
parton cascade. The unitarity constraint mentioned in the introduction
can be represented as \cite{GLR}
\beq
 W\,\,\leq\,\,1\,\,.
\eeq
Thus $W$ is indeed the small parameter sought. The parton cascade
can be expressed as a perturbation expansion in this parameter.
This perturbative series can in fact be resummed \cite{GLR} and
the result understood by considering the structure of the QCD
cascade in a fast hadron.

There are two elementary processes in the cascade that impact on
the number of partons.
\beq
{\rm splitting} \,\,\,\,(\, 1\,\rightarrow \,2\,);\,\,\, \,\,{\rm probability}
\,\,\propto\,\,\a_s \,\rho\,\,\,;
 \eeq
$$
{\rm annihilation}\,\,\,\,(\, 2\,\rightarrow\,1\,);\,\,\,
 \,\,{\rm probability}\,\,\propto\,\,\a^2_s\,
d^2\, \rho^2 \,\,\propto\,\,\a_s^2\,\frac{1}{Q^2}\, \rho^2\,\,,
$$
where $d$ is the size of the parton produced in the annihilation process.
In the case of deep-inelastic scatterinq $d^2 \sim 1/Q^2 $.
\par
When $x_B$ is not too small only the splitting of one parton into
two counts because $\rho$ is small. However as $x_B \rightarrow 0$
annihilation comes into play as $\rho$ increases.

This simple picture allows us to write an equation for the change
in the parton density in a `phase space' cell of volume
$\D\ln\frac{1}{x_B} \D\ln Q^2$:
\beq \label{EVPHYS}
\frac{\partial^2 \rho}{\partial \ln\frac{1}{x_B}\,\partial \ln Q^2}\,\,=
\frac{\alpha_s N_c}{\pi} \,\rho\,\,-\,\,\frac{\alpha^2_s
\,\gamma\pi}{Q^2}\,\rho^2\,\,,
\eeq
where $N_c$ is the number of colors.
In terms of the gluon structure function
\beq \label{GLR}
\frac{\partial^2 x_B G ( x_B,Q^2 )}{\partial \ln \frac{1}{x_B}\,\partial
\ln Q^2}\,\,=\,\,\frac{\alpha_s N_c}{\pi} x_B G (x_B, Q^2 )\,\,-\,\,
\frac{\alpha^2_s \,\gamma}{ Q^2 \,R_N^2} \,( x_B G (x_B,Q^2 ) )^2\,\,.
\eeq
This is the so-called GLR equation \cite{GLR}. To determine the
value of $\gamma$ and the understand the kinematical range
of validity of (\ref{GLR}) this simple physical description does
not suffice; rather one must analyze the process carefully
in $W$-perturbation
theory \cite{GLR} \cite{MUQI}.
The result for $\gamma$ was found to be
\cite{MUQI}
$$
\gamma\,\,\,=\,\,\,\frac{81}{16} \,\,\,\,\,\mbox{\rm for} \,\,N_c\,\,=\,\,3.
$$
We would like to emphasize that the main assumption in the above
derivation was that
\beq \label{RHOSQ}
 P^{(2)} \sim \rho^2,
\eeq
where $P^{(2)}$ denotes the probability for two gluons in
the parton cascade to have the same fraction of energy $x$ and
tranverse momentum (characterized by $r\simeq\ln Q^2$).

By assuming (\ref{RHOSQ}) we neglect all correlations between the
two gluons other than the fact that they are distributed in
the hadron disc of radius $R_N$. In the large $x_B$ region this assumption
is plausible because the correlations are power suppressed,
and the densities involved are small.
In the small $x_B$ region
we cannot justify (\ref{RHOSQ}), even when it holds for
large $x_B$.

\section{Induced multigluon correlations}

In \cite{TWOTW} it was shown that the problem is oversimplified
when one assumes that the probability of annihilation is simply
proportional to $\rho^2$ in deriving the GLR equation. It was found
that
\beq
\frac{P^{(2)}}{\rho^2} \,\,\propto\,\, e^{\frac{1}{ (N^2_c \,-\,1 )^2}\sqrt{
\frac{16 N_c \alpha_s}{\pi} \ln \frac{Q^2}{Q^2_0} \ln \frac{1}{x_B}}},
\eeq
where $Q_0$ is the initial virtuality in the parton cascade.
This ratio increases with decreasing $x_B$. Consequently we must take
dynamical correlations into account, which could change the GLR
equation crucially.

The key to the calculation of parton correlations was suggested
by Ellis, Furmanski and Petronzio in
\cite{ELPET}, and was developed further in \cite{BULI}. It was
shown that gluonic correlations are directly related to higher twist
contributions in the Wilson Operator Product Expansion (OPE)
arising from so called quasi partonic operators.
According to the OPE, the gluon structure function can be written as
\bea \label{HTSER}
x_B G (x_B,Q^2 )\,\, & =\,\,x_B G^{(1)} (x_B,Q^2)\,+\,
\frac{1}{Q^2 R_N^2} x^2_B
G^{(2)}(x_B.Q^2)\, \nonumber \\
& ...+\,... \frac{1}{(Q\,R_N)^{2(n - 1)}} x^{n}_B G^{(n)} (x_B,Q^2) \,...
\eea
where the $n$'th term results from the twist $2n$ quasi-partonic
operator. The probability density $P^{(n)}$ to find $n$ gluons in the cascade
with the same $x$ and $Q^2$ can be directly expressed through
the $n$'th term in the above expansion
($P^{(n)} = x_B^n G^{(n)}(x_B,Q^2)/(\pi R^2_N)^n, \;\, P^{(1)} = \rho$).

We recently determined the anomalous dimensions $\gamma_{2n}$
of these high twist operators \cite{HIGHTW} to next-to-leading
order in the number of colors $N_c$.
This was done by reducing the complicated
problem of gluon-gluon interactions to rescattering of
colorless gluon `ladders' (Pomerons) in the $t$-channel.
In \cite{TWOTW} it was shown that this approach works for
the case of the anomalous dimension of the twist 4 operator.
The fact that there is no Pomeron `creation' or `absorption'
in the $t$-channel means that we are dealing with a quantum
mechanical problem (not a field theoretical one) in the
calculation of the $\gamma_{2n}$ anomalous dimension. The problem
then amounts to calculating the ground state energy of an $n$-particle
system with an attractive interaction given by a four-particle
contact term (see Fig.1)
of strength $\lambda$. Its value can be calculated to be
\beq
 \lambda \,\,=\,\,4 \, \frac{\alpha_s N_c}{\pi} \,\delta,
\eeq
where $\delta = 1/(N_c^2-1)$ if one only takes color singlet ladders
into account. Including the other color states renormalizes $\delta$
to $0.098$ \cite{TWOTW}.
This effective theory is two-dimensional
(the two dimensions corresponding to $\ln(1/x)$ and $\ln(Q^2)$)
and is known as the Nonlinear Schrodinger Equation. It is well known
that this model is exactly solvable. Translating the ground state
energy of this model into the value of the anomalous dimension led
to \cite{HIGHTW}
\beq \label{ANDIM}
 \gamma_{2n} \,\,=\,\,\frac{\bar \alpha_s n^2}{\omega} \lbrace \,1\,\,+\,\,
\frac{\delta^2}{3} (\,n^2\,-\,1) \rbrace,
\eeq
where $\bar \alpha_s = \alpha_s N_c/\pi$,
$\omega = N-1$, $N$ being the Mellin-conjugate variable to
$x_B$
\beq \label{MELLIN}
f(N) = \int_0^1 dx_B x_B^{N-1}f(x_B)\;\; \mbox{{\rm or}}
\;\;f(\omega) = \int_0^\infty dy e^{-\omega y}[x_Bf(x_B)],
\eeq
where $f$ is an arbitrary function and $y=\ln(1/x_B)$.

The answer (\ref{ANDIM})
is only reliable when $\delta^2 n^2/3 \ll 1$ \cite{HIGHTW}.
Thus to check selfconsistency we must first generalize the GLR equation
based on (\ref{ANDIM}) and understand what value of $n$ is
important for the deep-inelastic structure function. If the
answer is inconsistent with the condition $\delta^2 n^2/3 \ll 1$
we must try and find the expression for the anomalous dimension
valid for any $n$.

Let us note that if we neglect the term proportional
$\delta^2$ in (\ref{ANDIM}) we can estimate  $x_B^n G^{(n)}(x_B,Q^2)$
via the inverse Mellin transform
\beq
\label{EIK}
x^n_B G^{(n)}(x_B,Q^2)\,\,=\,\,\frac{1}{2\pi i}\int_C d\omega
e^{(\omega y \,+\,\gamma_{2n} (\omega)r)} M^{(n)} (\omega,Q^2=Q^2_0)\,\,,
\eeq
where $y = \ln(1/x_B)$, $r = \ln(Q^2/Q^2_0)$, $Q_0$ being the initial
virtuality in the parton cascade.
The contour is to the right of all
singularities in M as well as to the right of the saddle point ($\omega_S$)
which is given by
\beq \label{SADDLE}
\frac{d}{d\omega}\{\omega y+\gamma_{2n}r\}|_{\omega = \omega_S}=0.
\eeq
Thus in the saddle point approximation for $\delta=0$
\beq
x^n_B G^{(n)}(x_B,Q^2) \,\sim \,[\, x_B G (x_B, Q^2)\,]^{n},
\eeq
in particular $P^{(2)} = \rho^2$.
We also assume here the factorization of the matrix
element $M^{(n)} \,=\,(\,M^{(1)}\,)^n$; this expresses the physical
assumption that there are no correlations between gluons other than the fact
that they are distributed in a disc of radius $R_N$.
We will comment more about this assumption further on.
In this sense the GLR
equation is only the lowest order approximation in $\delta^2$ even
if we assume the factorization of the matrix element and
thus, strictly speaking, is valid only in the limit of a large number
of colors. The term proportional to $\delta^2$ in
(\ref{ANDIM}) is clearly responsible for induced gluon correlations
and we take it seriously in this paper. In next section
we will therefore generalize the GLR equation.

\section{A New Evolution Equation}

The first step in such a generalization is to make an {\it ansatz}
for $P^{(n)}$ using the same approach as for the GLR equation,
viz. the competition of two processes in the parton cascade. Thus,
in analogy to the derivation of (\ref{EVPHYS}) we write
\beq \label{GENGLR}
\frac{\partial^2 \ P^{(n)}(x_B,Q^2)}{\partial \ln\frac{1}{x_B}
\,\partial \ln Q^2}\,\,=
 \,\,C_{2n} \cdot\,P^{(n)}(x_B,Q^2)\,\,-\,\,n\,\cdot\frac{\alpha^2_s
\,\gamma \pi}{Q^2}\,P^{(n + 1)}(x_B,Q^2)\,\,,
\eeq
where $C_{2n}=\gamma_{2n}\,\omega$.
The factor $n$ in front of the second term on the right hand side
of (\ref{GENGLR}) reflects the fact that in the Born
approximation $n + 1$ gluons annihilate in $n$ gluons through the
subprocess $ {\it gluon + gluon \,\rightarrow gluon}$,
which corresponds to the $two-ladder \rightarrow one-ladder$
transition with the strength of the triple ladder vertex $\gamma$.
There are $n$
such possibilities due to the time ordering of gluon emission.
Note that the infinitely recursive set of equations can be cut off
at any level by imposing e.g. $P^{(n)} = P^{(n-m)} P^{(m)}$. The GLR equation
is the case $P^{(2)} = (P^{(1)})^2$.
Since we operate under the assumption that high twists are essential
for small enough $x_B$ we must however consider the whole series in
(\ref{HTSER}), which we now do using eqs. (\ref{GENGLR}).

Let us introduce the generating function
\beq \label{GEXP}
g(x_B,Q^2, \eta )\,\,=\,\,\sum^{\infty}_{ n = 1} e^{n \eta} g^{(n)}\,\,,
\eeq
where $g^{(n)}=x^n_B G^{(n)}(x_B,Q^2)$. Comparing with (\ref{HTSER})
we see that for the full structure function
\beq \label{G}
x_B G(x_b,Q^2) = Q^2 R_N^2\, g (x_B,Q^2, \eta = - \ln (Q^2 R_N^2)).
\eeq
The recursive set of equations (\ref{GENGLR}) can be summarized in
one equation for $g$:
\beq \label{GENGLRPAR}
\frac{\partial^2 g (x_B,Q^2,\eta)}{\partial \ln \frac{1}{x_B} \partial \ln
Q^2 }=
\bar \alpha_s \frac{\partial^2 g}{\partial\eta^2}+\frac{\bar \alpha_s
\delta^2}{3} (\frac{\partial^4 g}{\partial\eta^4} -
\frac{\partial^2 g}{\partial\eta^2})
- \alpha_s^2 \gamma e^{- \ln (Q^2 R_N^2)} e^{ - \eta }( \frac{\partial
g}{\partial\eta}-g).
\eeq
To solve this linear, 4th order partial differential equation in
three variables, we must impose some boundary and initial conditions,
on the $Q^2$ and $x_B$ behavior respectively.

The boundary condition is straightforward
\beq
\mbox{{\rm For}}\;\; \eta,\; \ln(\frac{1}{x_B})\;\;
\mbox{{\rm fixed}}; \;\;\;
g (x_B,Q^2,\eta)\,\stackrel{\ln Q^2 \rightarrow \infty}{\rightarrow}
e^{\eta} \,g_{LLA}
(x_B,Q^2),
\eeq
where $g_{LLA} $ is the solution of the standard
GLAP evolution equation for the leading twist gluon density
in leading $\ln Q^2$ approximation.

The initial condition is much more difficult, because we need
$g(x_B = x_B^0, Q^2, \eta)$
for solving (\ref{GENGLRPAR}), whereas
experimentally we can only measure the structure function,
which is at fixed $\eta$.
In other words, we need detailed information on the gluon
distribution in a hadron at large $x_B$. We can make
the following suggestion (although others are possible)
\beq\label{EFOR}
g ( x_B^0,Q^2,\eta)=
\eeq
$$
\sum^{\infty}_{n = 1} e^{n \eta} \frac{ ( - 1
)^n}{n!} \cdot [\,g_{LLA} (x_B^0,Q^2)\,]^n \,=\, 1- \exp ( -
e^{\eta} g_{LLA} (x_B^0,Q^2)\,).
$$
This can be recognized as the usual eikonal approximation for
the virtual gluon-hadron interaction \cite{MUE}.
The virtues of this expression lie in the fact that it is simple,
and that it has the transparent physical meaning of reflecting
the assumption that there are no correlations between gluons
with $x\sim 1$ other than that they are distributed in a hadron
disc of radius $R_N$.

If one replaces the hadron with a nucleus, one can prove
such an approach, which corresponds to the so-called Glauber
Theory of shadowing corrections. In the deep-inelastic
scattering case an expression of this type was discussed by
A. Mueller in \cite{MUE}.

To summarize this section, we have proposed a new evolution
equation which has two new features over and above the GLR
equation:

(i) It includes induced multigluon correlations.

(ii) It allows an arbitrary initial condition not necessarily
an eikonal one, unlike the case of the GLR equation.
We recall that the GLR equation has been proven only under the assumption of
the
factorization property of the matrix elements,  which corresponds to an
eikonal initial condition \cite{GLR}.
By including all twists in our evolution equation we overcome
the need for a reductive initial condition, such as (\ref{EFOR}).
One could e.g. try to solve (\ref{GENGLRPAR}) using an initial
condition with correlated gluons at large $x$ and study the
consequences of its evolution, with or without $\delta$.
Because such initial correlations must be very small, and because
our main interest lies in comparing with the GLR equation, we
do not pursue this line of inquiry here.

\section{Approximate Solutions}

In the next section we will discuss the general solution to \eq{GENGLR}.
Here we  will give approximate solutions for various
special cases. In particular we try to answer the following
questions: how does the nonlinear GLR equation follow from
our present linear equation; what value of $n$ is typically
relevant in the sum (\ref{GEXP}) and how do the corrections
to the ordinary GLAP evolution due to (\ref{GENGLR}) differ
from those due to the GLR equation?

\subsection{The GLR Equation from the Generalized Equation}

The first question that arises is how the nonlinear GLR equation
is contained in the linear equation (\ref{GENGLR}) if we
neglect the term proportional to $\delta^2$. Specifically, we
would like to establish the equivalence of
\beq \label{SIMPLEEX}
\frac{\partial^2 g (x_B,Q^2, \eta )}{\partial \ln \frac{1}{x_B} \partial \ln
  Q^2 }\,\,=\,\,\bar \alpha_s \frac{\partial^2 g}{\partial\eta^2}
-\alpha_s^2\gamma \,e^{- \ln (Q^2 R_N^2)} e^{ - \eta }(\frac{\partial
g}{\partial\eta}-g),
\eeq
to the nonlinear GLR equation. Let us parametrize the solution in
the form
\beq
g (x_B,Q^2,\eta)\,\,=\,\,
\Phi(\,\,e^{\eta} F(y=-\ln x_B, r = \ln(Q^2 R_N^2)\,\,))\,\,.
\eeq
Because we dropped the gluon-correlation term from (\ref{GENGLR})
we can impose the `no-correlation' initial condition at
$x_B \sim 1$. This is the only initial condition the GLR equation
allows \cite{GLR}.
In the set of ``fan'' diagrams (Fig.3) that the GLR equation
sums the initial distribution has the form
\beq \label{FANIN}
g(x_{B}^0,Q^2,\eta)\,\,=\,\,\sum^{\infty}_{n = 1} e^{n\eta}( - 1 )^n \cdot
[\,g_{LLA}
(x_{B}^0,Q^2)\,]^n\,\,,
\eeq
which gives
\beq \label{FIFUN}
 \Phi (t) \,=\,\frac{t}{1\,+\,t}\,\,.
\eeq
The absence of the $1/n!$ in the above equation compared with (\ref{EFOR})
is explained as follows. The $1/n!$ in (\ref{EFOR}) enforces the
correct time ordering of the produced partons in the parton cascade,
related to diagrams of production of $n$ parton shadows
(see Fig.4 which shows a case in which three parton shadows are
produced). In the fan diagram of Fig.3 we do not have to enforce
the correct time ordering because it is already included
via the vertex $\gamma$ \footnote{The strength of the three Pomeron vertex
$\gamma$
was calculated in ref. \cite{MUQI} using the AGK cutting rules \cite{AGK},
which are equivalent to time-ordering.}.
Thus the initial condition of \eq{FIFUN} just corresponds to the sum of ``fan"
diagrams, of which Fig.3 is the lowest order example,
with the assumption that there are no correlations
between gluons inside the proton. We will see in section 6 that the
main properties of the full solution of \eq{GENGLR} do not depend on the
form of the initial condition. The reduction to the GLR equation does however.

With (\ref{FIFUN}) it is now straightforward to check that \eq{SIMPLEEX}
reduces to the GLR equation to first order in $e^{\eta}$ (recall that
finally we must put $\eta = -\ln(Q^2 R_N^2) \ll 1$), with
$F(y,r) = x G(x,Q^2)$.

\subsection{Relevant Twists in the Solution}

Here we try to follow the recipe mentioned below eq.~(\ref{MELLIN}):
to see if our approach is consistent we must determine what values
of $n$ are relevant in the solution to (\ref{GENGLRPAR}). Should those
values not be consistent with $\delta^2 n^2/3 \ll 1$ then we must
generalize the expression (\ref{ANDIM}) for $\gamma_{2n}$ such
that it is valid for all $n$. If they are consistent we
can maintain (\ref{ANDIM}) and thus (\ref{GENGLRPAR}).

However, to determine the relevant $n$'s we need
the general solution to eq. (\ref{GENGLRPAR}), which is presented
in section 6. Here we will perform some rough estimates.

Let us first take a `worst case' scenario by letting the
term proportional to $\delta^2$ dominate, and
by dropping the first
term and last term on the RHS of (\ref{GENGLRPAR}):
\begin{equation} \label{DELTAEQ}
\frac{\pa^2 g (x_B,Q^2, \eta )}{\pa \ln \frac{1}{x_B} \pa \ln Q^2 }\,\,=
\,\,\,\frac{\bar \alpha_s \delta^2}{3} (\frac{\partial^4 g}{\partial\eta^4}
- \frac{\partial g}{\partial\eta^2})
\end{equation}
To solve this equation we perform a Laplace transform in the
variable $\eta$
\beq
\ti  g(x_B,Q^2,p) \equiv \int_{-\infty}^0 d\eta \,\, e^{p\eta}
g(x_B,Q^2,\eta).
\eeq
Then
\beq \label{LAPDELTA}
\frac{\pa^2 \ti g (y,r,p)}{\pa y \pa r}\,\,=\,\,\frac{\bar \a_s \delta^2}{3}
\cdot
p^2(p^2 - 1)\,\,\ti g (y,r,p),
\eeq
where $ y=\ln(1/x_B), \;\;r=\ln(Q^2/Q^2_0)$. The solution to
(\ref{LAPDELTA}) is the Bessel function
$I_0(2\sqrt{\frac{\bar\a_s\delta^2}{3}p^2(p^2 - 1)y r})$.
The most general solution is then
\beq \label{SOLU}
g (y,r,\eta)\,\,=\,\,\int_C \frac{d p}{2\pi i}\,\, e^{- p\eta} \phi(p)\,
I_0(\,\, 2 \sqrt{\frac{\bar \a_s \delta^2}{3}p^2 (p^2 - 1) y r }\,\,)\,\,,
\eeq
where the contour $C$ runs to the right of all singularities in $p$.
The function $\phi(p)$ is fixed
by imposing an initial condition (e.g. (\ref{EFOR}))
\beq
g( 0,r ,\eta)
=\frac{1}{2 \pi i}\int_C dp \,\, e^{-p\eta} \phi(p).
\eeq
One may write the solution in fact directly in terms of the
initial condition:
\beq \label{SLTN}
 g (y,r,\eta)\,\,=\,\, \int^0_{- \infty}
 d \eta' G (y,r,\eta - \eta') g( 0,r, \eta'),
\eeq
with the Green's function
\beq \label{GREENFUN}
 G (y,r,\eta - \eta')\,\,=\,\,\frac{1}{2\pi i} \int d p\,\,
 e^{- p (\,\eta\,-\,\eta'\,)}
I_0(\,\, 2 \sqrt{\frac{\bar \a_s \delta^2}{3}p^2 (p^2 - 1) y r }\,\,)\,\,.
\eeq
{}From eq. (\ref{GEXP}) we see that large typical $n$ corresponds
to small typical $\eta$, which in turn by (\ref{SOLU}) corresponds
to large typical $p$ (``$p_0$''). Thus we must find $\delta^2 p_0^2 \ll 1$ to
trust eq. (\ref{ANDIM}).

Let us investigate (\ref{SLTN}) to find the most relevant values
of $\eta'$. The function $g(0,r,\eta')$ falls down monotonously
for $\eta' \rightarrow -\infty$ (the first term in (\ref{GEXP})
dominates) where it behaves as $\exp(\eta')$.
Next, we can show that the function
$G(y,r,\eta-\eta')$ has a maximum at $\eta=\eta'$ whose width
is of the order of $({8\sqrt{\frac{\bar \a_s \delta^2}{3} y r}})^{(1/2)}$.
To see this, evaluate (\ref{GREENFUN}) using the asymptotic form for
$I_0(z) \sim e^z/\sqrt{2\pi z} (1+\ldots)$.
Eq. (\ref{GREENFUN}) then becomes (neglecting the unimportant non-exponential
prefactor) in the asymptotic form
\beq \label{SADDLEDELTA}
G (y,r,\eta - \eta')\,\,=\,\,\frac{1}{2\pi i} \int d p\,\,
e^{- p (\,\eta\,-\,\eta'\,) \,+\,
2 \sqrt{\frac{\bar \a_s \delta^2}{3}p^2 (p^2 - 1) y r }}\,\,.
\eeq
The equation that determines the saddle point $p_S$ here is
\beq
 - ( \eta \,-\,\eta') \,\,+\,\,\frac{2 ( 2 p^2_S-1)}{\sqrt{p^2_S - 1}}
 \sqrt{\frac{\bar \a_s \delta^2}{3}\cdot y r }\,\,=\,\,0
\eeq
or
\beq
2 \sqrt{p^2_S - 1} \,+\,\frac{1}{\sqrt{p^2_S - 1}} \,=
\,\frac{\eta\,-\,\eta'}{2\sqrt{\frac{\bar \a_s \delta^2}{3}\cdot y r }}\,\,.
\eeq
In the dangerous case that $p_S$ is large this leads to
\beq \label{TYPP}
p_S \,\,=\,
\,\frac{\eta\,-\,\eta'}{4\sqrt{\frac{\bar \a_s \delta^2}{3}\cdot y r }}\,\,.
\eeq
With (\ref{SADDLEDELTA}) this becomes
\beq
 G (y,r,\eta - \eta')\,\,=\,\,\left(
8\pi\sqrt{\frac{\bar \a_s \delta^2}{3}\cdot y r }\right)^{-1/2}
\cdot e^{-\,
\frac{ (\eta\,-\,\eta')^2}{
8\sqrt{\frac{\bar \a_s \delta^2}{3}\cdot y r }}}\,\,.
\eeq
Thus, in summary, either or both the regions $\eta\sim 0$ and
$\eta' \sim \eta$ give dominant contributions in (\ref{GREENFUN}).
For $\eta'\sim 0$ we have
$$ g(y,r,\eta) \sim
e^{-\,\frac{\eta^2}
{8\sqrt{\frac{\bar \a_s \delta^2}{3}\cdot y r }}}\,\,.$$
while for $\eta' \sim \eta$
$$ g(y,r,\eta) \propto
    e^{-|\eta|}\,\,.$$
The first of these can dominate the second only for a restricted
range in $\eta$, namely\beq \label{ETAEST}
|\eta|<
8\sqrt{\frac{\bar \a_s \delta^2}{3}\cdot y r },
\eeq
which implies, with (\ref{TYPP})
\beq
p_S \,\,\ll \,\,2\,\,.
\eeq
Outside of the range (\ref{ETAEST}) the typical value
of $(\eta\,-\,\eta')$ is at large $\eta$ of order
$(8\sqrt{\frac{\bar \a_s \delta^2}{3}\cdot y r})^{(1/2)}$,
yielding
\beq
p_S \,\,\sim\,\,\frac{1}
{(2\sqrt{\frac{\bar \a_s \delta^2}{3}\cdot y r})^{(1/2)}}\ll1\;\;
\mbox{{\rm for}}
\,\,\,\,\,\,\,\, yr\,\,\gg\,\,1.
\eeq
We conclude, based on the simplified model equation
(\ref{SIMPLEEX}) that typical values of $p$, and thus $n$, are
small, and therefore we should be able to use eq. (\ref{ANDIM}).
We will see that this conclusion holds when we consider
the full solution in section 6.

\subsection{Estimates for Corrections to the GLR Equation}

In this subsection we want to estimate the possible size of
corrections to the GLR equation. Recall that the GLR equation
sums the contributions of fan diagrams (Fig.3) while the
generalized equation includes more general graphs, which are
all of the type shown in Fig.5
at large $Q^2$.
To estimate the correction let us parametrize the full solution as
\beq
g(x_B,Q^2,\eta)=
\Phi(\,\,e^{\eta} F(y,r)\,\,)+ \Delta g (y,r,\eta)\,\,,
\eeq
where the first term is the solution to the GLR equation
found in subsection 5.1, and the second term is considered
to be a small perturbation. Equation (\ref{GENGLRPAR}) becomes
then for $\Delta g$
\beq \label{PERT}
\frac{\partial^2 \Delta g (y, r, \eta )}{\partial y \partial r  }\,\,
=\,\,\bar \alpha_s \frac{\partial^2 \Delta g}{\partial\eta^2}
\,+\,\frac{\bar \alpha_s
\delta^2}{3} (\, \frac{\partial^4\Phi}{\partial\eta^4} \, - \,
\frac{\partial^2\Phi}{\partial\eta^2} \,)
\,-\,\Phi''\,[\,F'_y\,F'_r\,\,-\,\, \bar \a_s F^2\,]e^{ 2 \eta}\,\,,
\eeq
where $\Phi'$ denotes $\partial\Phi/\partial t$. We neglected the
contribution of
 the  $\partial\Delta g/\partial \eta
- \Delta g$ term in \eq{PERT} since the coefficient
in front of this term contains an extra power of $\as$ which we treat as
a small parameter. The initial conditions for
$\Delta g(y,r,\eta)$ are $\Delta g(0,r,\eta)$=
$\Delta g(y,0,\eta)=0$, because we assume that all boundary conditions
have been fulfilled by $\Phi$.
The last term in
eq. (\ref{PERT}) can be neglected for two reasons: (i) at
$\eta=-\ln (Q^2 R_N^2)$ this term is suppressed and (ii) the difference
in brackets is small in both the semicassical and the EKL \cite{EKL}
approach. Substitution of the explicit form of $\Phi$ in
(\ref{FIFUN}) yields
\beq \label{PERTFIN}
\frac{\partial^2 \Delta g (y, r, \eta )}{\partial y \partial r  }\,\,
=\,\,\bar \alpha_s \frac{\partial^2\Delta g}{\partial\eta^2}
\,+\,\frac{\bar \alpha_s
\delta^2}{3} \cdot\frac{ 12 t^2 ( t - 1)}{(1 + t)^5}\,\,,
\eeq
where $t\,=\,e^{\eta} F(y,r) $.  We can simplify the equation if we
keep in mind that the value of $\eta$ is large and negative
in the deep-inelastic structure function. Thus $t\ll 1$ and
\beq \label{PERTSIMPLE}
\frac{\partial^2 \Delta g (y, r, \eta )}{\partial y \partial r  }\,\,
=\,\,\bar \alpha_s \frac{\partial^2\Delta g}{\partial\eta^2}
\,- 12\,\frac{\bar \alpha_s
\delta^2}{3} F^2 e^{2 \eta}\,\,.
\eeq
Now the $\eta$ depedence of $\Delta g$ is trivial:
$\Delta g = e^{2 \eta} \Delta F (y,r)$. Thus
\beq \label{PERTCOR}
\frac{\partial^2 \Delta F (y,r)}{\partial y \partial r  }\,\,
=\,\,4\,\bar \alpha_s \Delta F(y,r) \,- 12\,\frac{\bar \alpha_s
\delta^2}{3} F^2 \,\,.
\eeq
Using similar techniques as in the previous subsection, but now
for the $y$ and $r$ variables, it is straightforward to show that the
general solution to this equation with the initial condition
$\Delta F (0,r ) \,=\,0$ has the form
\beq \label{GLRDF}
\Delta F \,\,=\,\,
-\,12\frac{\bar \alpha_s \delta^2}{3}
 \int^y_0 d y'  \int \frac{d f}{2\pi i} \int \frac{d\omega}{2 \pi i}
e^{\omega y' \,+\,f r } \,\,\frac{\omega\,\tilde{  F^2}(\omega,f)}{
\omega f\,\,-\,\,4\,\bar \alpha_s}\,\,,
\eeq
where $\tilde{ F^2}$ is the Laplace transform in $y$ and $r$
of the function $F^2(y,r)$. The contours for the $f$ and $\omega$ integrals
lie to the right of all singularities in these variables.
We now need a reasonable estimate for
$\tilde{ F^2}$.
Note that  $\tilde{ F^2}$ corresponds to the (Laplace tranform of)
the properly normalized initial gluon distribution at
low virtuality and small $y$.
A rough estimate can be made usine the methods
of \cite{EKL}.
\beq \label{EKLANS}
\tilde{F^2}\,\,=\,\,\frac{A^2_G}{ [\,\omega \,-\,2\,\omega_0\,][\,f\,-\,2\,
\gamma
(\omega =\omega_0)]}\,\,,
\eeq
where $A_G$ is the normalization factor for the gluon structure function,
$\omega_0$ is defined in eq. (1) and
$\gamma(\omega)\,=\,\frac{\bar \a_s}{\omega}$ is the anomalous dimension for
the leading twist operator.
Here we consider the EKL solution to parametrize the data over a wide
kinematic region, including GLR nonlinear corrections. Shortly we
will discuss the case where both the GLR and multigluon corrections
are considered small.

All contours in (\ref{GLRDF}) are to the
right of all singularities in $ \omega$ and $f$.
The integrand has two poles in $\omega$, one corresponding to the
initial condition ($2\omega_0$) and one from
the equation ($4\bar \alpha_s/f$).
We will demonstrate in the next subsection that the former
is dominant for the choice $\omega_0 = 0.5$, and that restricting
ourselves to its contribution is very good approximation.
Under this assumption we perform the
$\omega$ and $f$ integrals, and obtain
\beq
\Delta g \,\,=\,\,-\,4 r  \bar \a_s \delta^2  A^2_G e^{2\eta} \int^y_0 d y'
 e^{2\omega_0\,y' \,+\,2\,
\gamma(\omega\,=\,\omega_0)\,r }\,\,.
\eeq
The factor $r$ in front arises from the double
pole in the $f$ variable. The answer clearly satisfies the boundary
conditions.
We derive further
\beq
\frac{\Delta g (y,r,\eta)}{g (y,r,\eta)}\,\,\simeq\,\,
- \frac{2 {\bar\a_s} r}{\omega_0} \delta^2 e^{\eta}\cdot
\frac{ F^2(y,r) \,-\,F^2( y = 0, r )}{F(y,r)}\,\,.
\eeq
This implies
\beq \label{FINEST}
\frac{\Delta x_B G (x_B,Q^2)}{ x_B G(x_B,Q^2)}\,\,=\,\,
-\,\frac{2 \bar{\a_s} r}{
 \omega_0} \delta^2 \cdot\frac{1}{Q^2R_N^2}   x_B G(x_B,Q^2)\,\,,
\eeq
if the value of the structure function is large enough in the region of
low $x_B$. Recall that the correct definition of $\eta$ is
$e^{\eta}\,= \,1/(Q^2 R_N^2)$.
Substituting in eq. (\ref{FINEST}) the value of gluon structure function
from HERA data \cite{HERA} at $Q^2\,=\, 15\; {\rm GeV}^2$ and $x_B = 10^{-4}$
($ x_B G \,\sim\, 30$) and a typical value of $R_N^2 = 5 GeV^{-2}$
we obtain
(with $\a_s\,=\,0.25$ )
$$ \frac{\Delta x_B G (x_B, Q^2 )}{ x_B G (x_B,Q^2)} \,\,\sim\,0. 4
  \delta^2 \,\sim\, 6\cdot 10^{ - 3}\,\,.$$

It is more instructive to compare the above correction
to the gluon structure function  with the one due to the GLR equation.
In this case we consider both
the correction to the GLAP equation due to the GLR shadowing and due to
multigluon correlations as small. Thus we
try to find the solution to the GLR equation in the form:
$$
F(y,r)\,\,=\,\,F_{GLAP} \,\,+\,\,\Delta F_{GLR}.
$$
For $\Delta F_{GLR}$ we can write the equation:
\beq
\frac{\pa^2 \Delta F_{GLR}}{\pa y \,\pa r}\,\,=\,\,\bar \as \,\Delta F_{GLR}
\,\,-\,\,\frac{\as^2 \gamma}{Q^2 R^2_N}\,F^2_{GLAP}
\eeq
The solution is
\beq
\Delta F_{GLR} \,\,=\,\,-\,\,\frac{\as^2 \gamma}{Q^2R^2_N} \,\int^y_0 d y' \int
\frac{d f}{2 \pi i}\,\int\frac{d \omega}{2 \pi i}\,\,e^{\omega y'\,+\, fr}
\,\frac{\omega \tilde{F}^2_{GLAP} (\omega, f )}{ \omega f\,-\,\bar \as},
\eeq
where $\tilde{F}^2_{GLAP}$ is the Laplace image of function $F^2_{GLAP} (y,r)$.
Again using (\ref{EKLANS}) we get
\beq
\Delta F_{GLR}\,\,=\,\,-\,\frac{2\omega_0 \as^2 \gamma}{3\bar\as Q^2 R^2_N}
\,A^2_G \,
\int^y_0 dy'
e^{2 \omega_0 y'}\,[ e^{2 \gamma( \omega = \omega_0 )}-e^{ \gamma(
 \omega = \omega_0 )}]
\eeq
Assuming that the second term in the above equation is much smaller than the
first we have
\beq
\Delta F_{GLR}\,\,=\,\,-\,\frac{\as \gamma\pi}{3 Q^2 R^2_N N_c}\,F^2_{GLAP}
\eeq
Finally, we can get  for the ratio
\beq
\frac{\Delta x_B G (x_B, Q^2)}{ \Delta (x_B G( x_B,Q^2))_{GLR}}\,\,
=\,\,\frac{6 N_c^2 \ln (Q^2/Q^2_0)}{\pi^2 \omega_0 \gamma } \,\delta^2\,\,
\eeq
which gives a value of the order of 0.04 if  $\omega_0 \sim 0.5$.
We will return to a discussion of the corrections
to the GLR equation in the next section where we will consider the general
solution to the new equation at fixed $\as$.

The above estimates seem in
contradiction with the estimates in \cite{BARY}, where a large contribution
from multigluon correlations was found.
The method we use here is however quite diffent from the one in \cite{BARY}.
There the effect of including the new (pole) singularity in
$f$ (the Laplace conjugate variable to $r$) that results
from the resummation of bubbles associated with the 4-Pomeron
coupling was contrasted with the contribution from the two Pomeron cut at
the level of the Green functions. Although the location of the singularities
is quite close, their nature and residues are very different.
This led to a large ratio of the contributions of these singularities
to the Green functions.
In the present case we include in our estimates the two-gluon source,
i.e. the initial gluon distribution, and renormalize it in both cases
to the same physical initial condition
(here the EKL ansatz), thus absorbing the residues.
As was remarked in \cite{BARY}, one can absorb the residues alternatively
in $R_N$.
Further, in closing the contours involved in performing the inverse Laplace
transforms we closed on the singularities of
the initial condition (being the rightmost singularities),
and not on the propagator poles. Therefore our renormalized $R^2_N$
has no extra dependence on $\ln(1/x)$. We believe
that our method is in the above sense more physical.

\subsection{Numerical Estimates}

In this subsection we will solve eq. (\ref{PERTCOR}) numerically.
The method we use goes as follows. We first perform a Laplace transform
with respect to $y$ on (\ref{PERTCOR}), which leads to
\beq \label{GLRDFOM}
\frac{\partial \D F(\omega,r)}{\partial r} =
\frac{4{\bar \as}}{\omega} \D F(\omega,r) -
\frac{4{\bar \as}\delta^2}{\omega} F^2(\omega,r)\,\,.
\eeq
Using various {\it ans\"{a}tze} for $F(y,r)$ we solve this
equation using Numerov's method, and perform finally the
inverse Laplace transform with respect to $\omega$ \footnote{We thank
Keith Ellis for providing subroutines that perform this last step.}.
For the various {\it ans\"{a}tze} we fit a parametrized form
to $F^2(y,r)$ of which the Laplace transform can be taken
analytically.

We will investigate three cases. The first is the EKL ansatz
from the previous subsection
\beq \label{EKL2}
F(y,r) = A_G \,e^{\omega_0 y + f_0 r} \,\,,
\eeq
where we take $\omega_0 = 0.5$, $f_0 = {\bar \as}/\omega_0$,
and ${\bar \as} \simeq 0.25$. To correspond with the
numbers given in the previous subsection we put
$A_g \simeq 0.07$.
We will use this case to check the accuracy of the estimate given
earlier,
and drop for this check the requirement $\D F(y=0,r) = 0$.
Let us denote the numerical answer by $\D F^{\rm EKL}(num)$.
Then for the pole answer we find
\beq \label{EKLPOLE}
\D F^{\rm EKL}(pole) = \frac{-2{\bar \as}\delta^2}{\omega_0}
\,r\, e^{2\omega_0 y + 2\gamma_0 r}\,\,,
\eeq
and for the leading term of the saddle point contribution
\beq \label{EKLSADDLE}
\D F^{\rm EKL}(saddle) \simeq \frac{2{\bar \as}\delta^2}{\omega_0}
\sqrt{\frac{f_S^3}{16\pi {\bar\as}y}} \frac{1}{(f_S-2\gamma_0)^2}
e^{4\sqrt{{\bar\as} y r}}
\eeq
where $f_S = \sqrt{4{\bar\as}y/r}$. Note that both contributions
vanish as $r\ra 0$.
In Table 1 we list these three contributions for various
$x$ at a fixed $Q^2 = 15$ ${\rm GeV}/c^2$.
It is clear from this table that
$\D F^{\rm EKL}(pole)$ is a very good approximation
to $\D F^{\rm EKL}(num)$, in fact better than one might expect
from $\D F^{\rm EKL}(saddle)$.

Next we treat a more realistic case. We now use for $F(y,r)$
alternatively the MRSD0' and MRSD-' \cite{MRS} gluon distribution
functions, which are, at the starting scale $Q_0$, constant
as function of $x$, and behave as $x^{-0.5}$ respectively.
They are both parametrized by
\beq \label{MRSPAR}
x G(x,Q^2) = A_G x^{\lambda_g} (1-x)^{\eta_g}
(1+\gamma_g x)\,\,,
\eeq
with the coefficients $A_G, \lambda_g, \eta_g$ and $\gamma_g$
given at $Q=Q_0$ in \cite{MRS}.
We kept this parametrization up until $Q^2 = 15$ ${\rm GeV}/c^2$,
but refitted the coefficients for every step in the Numerov
procedure.
The Laplace transform of this parametrization is easily
determined. Following the methods described in the above
we determine $\D F(y,r)$. We checked that $\D F$
vanishes for small $y$. Furthermore \eq{GLRDFOM} was
solved under the condition $\D F(y,r=0) = 0$.

Similarly to the previous subsection we determined
at $Q^2 = 15$ ${\rm GeV}/c^2$ for values of
$x_B$ from 0.01 down to the ${\rm LEP}\otimes{\rm LHC}$
value of $10^{-5}$ the ratios
\beq \label{RAT1}
\frac{\D (x_B G(x_B,Q^2))}{x_B G(x_B,Q^2)}
= \frac{1}{Q^2 R_N^2} \frac{\D F(y,r)}{x_B G(x_B,Q^2)}
\eeq
and
\beq \label{RAT2}
\frac{\D (x_B G(x_B,Q^2))}{\D(x_B G(x_B,Q^2))_{GLR}}
= \frac{\D F(y,r)}{\as^2 \gamma (x_B G(x_B,Q^2))^2}.
\eeq
These ratios are given in Table 2. We infer from this table
that corrections to the gluon structure function from multigluon
correlations beyond the next-to-leading twist are small,
at most 5\% at small $x$. We see that the MRSD-' distribution
leads to larger corrections than the MRSD0' one.
As a fraction of the GLR correction both {\it ans\"{a}tze}
are small, about 10\% for the MRSD-' case and
10-30\% for the MRSD0' case.

Thus we confirm here the rough estimates from the previous
section. However we note that estimates of the gluon
correlation radius $R_N$ range from  $1$ fm to $0.3$ fm.
The numbers in table 2 are accordingly easily adjusted.

\section{The General Solution (for Fixed $\as$)}

In this section we will discuss the general solution to \eq{GENGLR}
and its consequences for the gluon structure function.

We start by noting that the equation (\ref{GENGLR}) can be written, at
fixed $\as$, as
\bea \label{XIGEN}
\frac{\pa^2 x_B^n G^{(n)}( x_B, r + \eta_0)}{\pa y \pa ( r +
\eta_0)}=
\,C_{2n}\,\cdot  x_B^n G^{(n)} ( x_B, r + \eta_0)
&-&\\n\cdot\,\as^2 \gamma e^{-(r+\eta_0)}& x_B^{(n+1)}
G^{(n + 1)} ( y, r + \eta_0),& \nonumber
\eea
where we used the fact that
that $x_B^nG^{(n)}$  only depends on $\ln(Q^2/Q^2_0)$
with an arbitrary $Q_0$.
The above form of the equation reflects the choice
$\eta_0 = -\ln Q^2_0 R_N^2$.
We focus now on the hypersurface $\eta_0 = \eta$.
The evolution equation (see \eq{GENGLRPAR} ) can then be written in the form:
\beq \label{GENGLRPARXI}
\frac{\partial^2 g (y,\xi,\eta)}{\partial y \partial
\xi}=\bar \alpha_s \frac{\partial^2 g}{\partial\eta^2}+\frac{\bar \alpha_s
\delta^2}{3} (\frac{\partial^4 g}{\partial\eta^4} -
\frac{\partial^2 g}{\partial\eta^2})
- \alpha_s^2 \,\,\gamma \,\,e^{- \xi }( \frac{\partial
g}{\partial\eta}-g)\,\,,
\eeq
where $\xi \,=\,r\,+\,\eta$. The advantage of this form is that all explicit
$\eta$ dependence such as $\exp(-\eta)$ has been removed.

We can find $ g (y, \xi = r + \eta, \eta )$ using a double Laplace transform
with respect to $y$ and
$\eta$, namely
\beq \label{DOUBLEMEL}
g (y,\xi,\eta ) \,\,=\,\,\int
\frac{d \omega\,\,d \,p}{ ( 2\pi i)^2}\,\,
e^{\,\omega\,y\,\,+\,\,p\,\eta} \,\,
g ( \omega, p , \xi ) \,\,.
\eeq
The function $g (\omega,\xi,p)$ obeys the equation
\beq \label{XIEQ}
\omega\,\frac{\pa g (\omega, \xi,p)}{\pa \xi}\,\,=
\{ \,\bar \as
\,p^2\,+\,\frac{\bar \as \delta^2}{3}\,(p^4-p^2)\,-\,
\as^2 \gamma (p\,-\,1)\,e^{-\,\xi}\,\}\,g (\omega, \xi,p)
\eeq
The solution to \eq{XIEQ} is:
\beq \label{SOLXI}
g (\omega, \xi,p )\,=
\,\int
\frac{d\,\omega\,d\,p}{(2\pi i)^2}\,\,g (
\omega, p )\cdot e^{(\,\frac{ \bar \as }
{\omega}\,p^2\,\,+\,\, \frac{{\bar \as} \delta^2}{3\,\omega}\,p^2 ( p +
1)( p - 1 )\,)\,\xi \,\,+\,\,\frac{{\bar \as}^2 \,\gamma}{\omega}( p - 1)
\,( e^{-\,\xi}\,-\,1)}
\eeq
The function $g (\omega, p )$ must be determined from the initial
condition, \eq{EFOR},
for $ | \eta | \,\ll\,r$, $ r\,\gg\,1$ and $ y = 0 $ where the solution
looks as follows:
\beq \label{INCON}
g (y = 0, r, \eta)\,=
\,\int \frac{ d \omega d p}{ ( 2 \pi i )^2}
\,g(\omega,p)\,\cdot\,e^{ p\,\eta}
e^{ ( r + \eta) \,\phi (p )\,\,+\,\,\frac{\as^2 \,\gamma}{ 3
\,\omega} \,(\, p\,-\,1\,) \,(\,e^{-r\,-\,\eta}\,\,-\,\,1\,)}
\,,
\eeq
where
$$
\phi(p)\,\,=\,\,\frac{\bar \as}{\omega}\,p^2\,+\,\frac{{\bar
\as}\,\delta^2}{3 \omega} \,p^2\,( p + 1)( p - 1 )\,\,.
$$
We now assert that
\beq \label{OMP}
g (\omega, p )\,\,=\,\,\Gamma (-\, f (p))\,\cdot\,\frac{d\,f(p)}{d\,p}\,
\,\,e^{\frac{\as^2\,\gamma}{\omega}}
\eeq
satisfies the initial condition of \eq{EFOR} with $g_{LLA}
(x_{B},Q^2_0)\,=\,\delta ( y ) $ at $r\,=\,\ln(Q^2/Q^2_0)$ = 0.
Here $ f(p)\,\,=\,\,p\,\,+\,\,\phi
( p ) $ and $\Gamma (- f )$ is the Euler gamma function.
We will prove this shortly.

Thus, finally, the solution to \eq{GENGLRPARXI} looks as follows:
\beq \label{FINSOL}
g (y, \xi, \eta
)\,=
\,\int\frac{d\,\omega\,d\,p}{(2\pi i)^2}\,\,
\Gamma (- f (p))\,\cdot\,\frac{d\,f(p)}{d\,p}\,
\,\cdot
 e^{f(p)\,\xi \,\,+\,\,\frac{\as^2
\,\gamma}{\omega}\,( p   - 1) \, e^{-\,\xi}\,\,-\,\,p\,r\,\,+\,\,\omega\,y}
\eeq
or changing the integration from $p$ to $f$,
\beq \label{FINSOLF}
g (\omega, p, \xi
)\,=\,\int\,\frac{d\,\omega\,d\,f}{(2\pi i)^2}\,\,
\Gamma (-\, f )\,
\cdot\,
 e^{\,f\,\xi \,\,+\,\,\frac{\as^2
\,\gamma}{\omega}\,( p(f)   - 1) \, e^{-\,\xi}\,\,-\,\,p(f)\,r
\,\,+\,\,\omega\,y}\,\,,
\eeq
where $p (f)$ is determined by
\beq \label{FP}
f\,\,=\,\,p (f)\,\,+\,\,\phi( p(f) )\,\,.
\eeq
The contour of integration over $f$ is defined in a such a way that all
singularities in $\Gamma( - f ) $ except the one at $f=0$
are located to the right of the contour.

We can now verify the claim made in \eq{OMP}. At $y=y_0$,
$|\eta| \ll r$, $\xi\simeq r \gg 1$ we have
\beq\label{CHECKINIT1}
g(y_0,r,\eta) \simeq \int \frac{d\omega \, df}{(2\pi i)^2}\,
e^{\omega y_0} \Gamma(-f)\,
e^{-r p(f) + \xi f}.
\eeq
We now close the $f$ contour in the right half plane.
Neglecting the $\delta^2$ term we have $\phi(p(n)) \simeq \bar\as n^2/\omega$,
and
we get indeed
\beq\label{CHECKINIT2}
g(y_0,r,\eta) \simeq \sum_{n=1}^{\infty} \frac{(-)^n}{n!}
e^{n\eta} \int \frac{d\omega}{2\pi i}
e^{\omega y_0 + \frac{\bar \as n^2}{\omega}}
\eeq
The structure function can be found from \eq{FINSOLF} putting $\xi=0$
(see \eq{G}). We obtain
\beq \label{STRSOL}
 g(y, \xi=0,r)
\,\,=\,\,\int\,\frac{d\,\omega\,d\,f}{(\,2\pi\,i\,)^2}\,\,
\Gamma (-\, f )\,\cdot
\, e^{\frac{\as^2
\,\gamma}{\omega}\,( p(f)   - 1) \,\,-\,\,p(f)\,r
\,\,+\,\,\omega\,y}\,\,.
\eeq
\Eq{FP} implies that
$p(f) \,\ra\, (\,\frac{3\,\omega\,f}{\bar \as \delta^2}\,)^{\frac{1}{4}}$.
Therefore we can close the contour
in $f$ over the singularities of $\Gamma ( - f)$.
Next we must integrate over $\omega$. We evaluate this integral
using the method of steepest descent and the large $f$ approximation
for $p(f)$. Neglecting terms proportional to
$\as$ in the exponent we find the saddlepoint
$\omega = (p_0 r/4 y)^{4/3}$ where $p_0 = (3n/{\bar\as}\delta^2)^{1/4}$.
Then
$$
g ( y, \xi = 0, r ) \,\,=\,\,\sum_{n=1}^{\infty} \,\,\frac{
(\,-\,1\,)^n}{n!}\,C (n,y)
e^{-\,4^{-\frac{1}{3}}\frac{3}{4 }\,(\,\frac{3 \,n}{\bar \as
\delta^2}\,)^{\frac{1}{3}}\cdot (\,\frac{r^4}{y}\,)^{\frac{1}{3}}}\,\,,
$$
where $C (n, y)$ is a pre-exponential, smooth factor. The above series clearly
converges and the typical value of $n$ in this series is of
the order of unity. We thus confirm the estimates from the previous
section, and
see that we could trust our calculations of the
anomalous dimension $\gamma_{2n}$.
We note that this series has an infinite radius of convergence.
Thus for the simplified model we consider and for the case of
an eikonal initial condition we conclude that there is
no analogy to a renormalon in the Wilson Operator Product Expansion.

Now we wish to consider the solution near $f\,\ra\,1$,
to understand under which circumstances it suffices to take
only this singularity into account.
Note that this corresponds to the leading twist case, with
GLAP evolution.
In this region we can
rewrite the solution in the following form.
Substituting
$f \,=\,1 +\,t$ we obtain:
\beq
\label{SOLF1}
 g(y, \xi = 0,r )\,\,=\,\,\int \frac{d \omega}{2 \pi i}\,\frac{d
t}{2 \pi i } \,\frac{1}{t}\cdot e^{\Psi}
\eeq
where
\beq \label{PSI}
\Psi\,\,=\,\,\omega\,y\,\,-\,\,r\,\,+\frac{\bar
\as}{\omega}\,r\,\,+\,\,t\,\,r\,\{\,\frac{2 \bar \as
(\,1\,+\,\frac{\delta^2}{3}\,)}{\omega}\,\,+\,\,\frac{ \as^2
\,\gamma}{\omega\,r}\,\,-\,\,1\,\}
\eeq
{}From this form of the exponent we see
that for $\omega$ smaller than $\omega_{cr}$ where
\beq \label{OMCR}
\omega_{cr} \,\,=\,\,2\,\bar \as (\,1\,+\,\frac{\delta^2}{3}\,)\cdot\{\,1
\,+\,\frac{\as^2\,\gamma}{2{\bar\as}(1+\delta^2/3)\,r}\,\}
\equiv\omega^0_{cr}\,\{
\,1\,+\,\frac{\as^2\,\gamma}{\omega^0_{cr}\,r} \,\}
\eeq
we cannot close the contour in $t$ over singularities at positive $t$.
(Here we introduce the parameter $\omega^0_{cr}$ in order to separate
the effects of $\gamma$ and $\delta$.)
For such $\omega$ one would need all singularities in $f$.
To
understand what happens in this region let us expand $\Psi$ by writing
$\omega\,\,=\,\,\omega_{cr}\,\,+\,\,\D$:
\beq \label{PSI2}
\Psi\,\,=\,\,\omega_{cr}\,y\,\,-\,\,r\,\,+\,\,\frac{\bar
\as}{\omega_{cr}}\,r\,\,+\,\,\D\,(\,y \,\,-\,\,\frac{\bar
\as}{\omega^2_{cr}}\,r
\,\,-\,\,\frac{t\,\,r}{\omega_{cr}}\,)
\eeq
Integration over $\D$ \footnote{Alternatively one may expand to
$O(\D^2)$ and use steepest descent.}
gives rise to the delta function
$\delta [\, \frac{r}{\omega_{cr}}\,(t\,\,-\,\,t_0\,)\,]$
where
\beq \label{TO}
t_0\,\,=\,\,\frac{1}{\omega_{cr}\,\,r}\,[\,\omega^2_{cr}\,y-\bar \as\,r]
\eeq
Carrying out the integration gives
\beq \label{SOLFINF1}
 g(y, \xi = 0,r)
\,\,=\,\,\frac{\omega_{cr}}{2\pi r}
\,\cdot\,\frac{1}{t_0}\,\cdot\,e^{\omega_{cr}\,y\,\,-
\,\,r\,\,+\,\,\frac{\bar \as}{\omega_{cr}}\,r}
\eeq
Let us define the ``critical line'' \cite{GLR} by
\beq \label{CRLINE}
y_{cr}\,\,=\,\,\frac{1}{\omega^0_{cr}}\,r\,\,-\,\,\frac{\bar
\as}{(\,\omega^0_{cr}\,)^2}\,r
\eeq
The gluon structure function is on this critical line
\beq \label{STRFUNCRLINE}
x_B G( x_B, Q^2 )=R_N^2\,Q^2\,\,\frac{(\,\omega^0_{cr}\,)^2}{2\, {\bar
\as} r} \,[\frac{\delta^2}{3}\,\,+\,\,\frac{\as^2\gamma}
{4\pi\bar \as \,r}\,]^{-1}
\eeq
For $\delta\,=\,0$ this becomes
\beq \label{STRFUNDO}
x_B G( x_B, Q^2 )=R_N^2\,Q^2\,\,\frac{2}{\gamma} \frac{N_c^2}{\pi^3}.
\eeq
Note that this is similar to the solution of the GLR equation with running
coupling \cite{GLR}, but not quite the same.

Thus, the  structure of the solution to the new evolution equation
with  the initial condition \eq{EFOR}
looks as follows. In the kinematic region
to the right of the critical line we can in fact
solve the linear
GLAP equation, but with the
new initial condition \eq{STRFUNDO}  on the critical
line.
To the left of the critical line we need the solution to
the full equation.
Note that if we would change the initial condition
of the full evolution equation the value of the structure
function on the critical line would also change.

We have thus achieved a
new understanding of the role of the initial condition in the problem. In
particular, we conclude that the solution on the critical line
depends only on the
initial condition in the region of $f\,\ra \,1$, i.e.
it depends only on the initial condition for the GLAP equation.
Implicitly we have used here the assumption, expressed in
our initial condition (\ref{EFOR}), that the multigluon
correlations are sufficiently small at large $x$.
We recall
that the original derivation of the GLR equation was based on this
same assumption.
We are not restricted to such an assumption for our
equation. Clearly, if there were strong correlations between gluons
at large $x$
it would change the explicit form of the solution of our
evolution equation. Nevetheless, the line of reasoning followed in this
section would continue to hold.

For $\delta\,\neq\,0$ we have two different situations.
In the first, for
$$
\frac{\delta^2}{3}\,\,\ll\,\,\frac{\as^2\,\gamma}{2 \bar \as \,r}
$$
the solution on the critical line is the same as in \eq{STRFUNDO}.
I.e. the only change that occurs in the solution of
the case $\delta\,=\,0$ is that there is a new equation for
the critical line,
\eq{CRLINE}.
Note that the HERA experiments
correspond to this situation.

At very large values of $r$ ($Q^2 \,\gg\,Q^2_0$) , when
$$
\frac{\delta^2}{3}\,\,\gg\,\,\frac{\as^2\,\gamma}{2 \bar \as \,r},
$$
the solution on the critical line looks as follows:
\beq \label{STRFUNDNEQO}
x_B G( x_B, Q^2 )=R_N^2\,Q^2\,\,\frac{3\,\omega^0_{cr}}
{2\pi\delta^2\,r}
\eeq
In this case we must solve the GLAP equation using \eq{STRFUNDNEQO} as the
boundary condition.

\section{Conclusions.}

The main result of the paper is the new evolution equation
(\ref{GENGLRPAR}). It allows us
to penetrate deeper
into the region of high density QCD because we incorporate multigluon
correlations into the evolution,
and to answer questions which could not be answered before.
For example one could now investigate
the question of how well the Glauber theory
for shadowing corrections in deep-inelastic scattering with a
heavy nucleus works.

This present equation solves two
theoretical problems which arise in the region of high parton density:
(i) It takes  induced multigluon correlations into account,
which originate from parton-parton (mainly gluon-gluon)
interactions at high enegy
and can be calculated in the framework of perturbative QCD,
and (ii) it allows for an arbitrary initial gluon distribution,
which is nonperturbative in nature.

We have found the general solution to the new equation for the
case of an eikonal initial condition and fixed $\alpha_s$. We
found no evidence for a ``renormalon'' in the twist expansion.

Our numerical estimates show that the effect of multigluon correlations is
rather small in the accessible region of energy.
We have seen evidence for this
by using approximate methods
and the general solution to the new equation.

We have shown that the general solution
confirms the strategy developed for the GLR
equation:
we have calculated the new critical line for the generalized
equation and shown that to the right of this critical line we can solve the
linear GLAP equation with a new boundary condition on this line.
We found this boundary condition
taking into account the multigluon
correlations. This approach,
developed in this paper simplifies also the solution
to the GLR equation and allows us to understand how solutions to the GLR
equation depend on the initial conditions. This is essentially
a consequence of the linearization of the GLR equation
in \ref{GENGLRPAR}.

We have not discussed here the behaviour of the solution in the
region to the left of the critical line, where multigluon correlations
should come more forcefully into play.
We plan to do this in later
publication.
We hope that the  solution in the latter kinematic region will have a
significant impact on understanding the scale of the shadowing correction
and the importance of multigluon correlations in the so-called Regge domain.
This must be understood in order to provide a matching between soft and hard
processes.

\noindent
{\bf Acknowledgments}
E. Laenen wishes to thank the Columbia University theory
group for their hospitality. E. Levin acknowledges the financial support
by the Mortimer and Raymond Sackler Institute of Advanced Studies and by the
CNPq.


%

\newpage

\begin{tabular}{||c||c|c|c||} \hline
 $x_B$ & $\Delta F^{\rm EKL}(num)$ & $\Delta F^{\rm EKL}(pole)$
& $\Delta F^{\rm EKL}(saddle)$ \\ \hline
$10^{-2}$  & $-3.26\cdot 10^{-2}$   & $-3.41\cdot 10^{-2}$ &
       $8\cdot 10^{-3}$  \\
$10^{-4}$  & $-3.44$  & $-3.41$  & $2\cdot 10^{-2}$  \\ \hline
\end{tabular}
\vspace{0.4cm}

Table 1. Comparison of various contributions in the EKL
approximation.
Here $Q^2 = 15$ ${\rm GeV}^2/c^2$, $\alpha_s = 0.25$
and $R_N^2 = 5$ ${\rm GeV}^{-2}$.

\vspace{4cm}
\begin{tabular}{|c||c|c|c||} \hline
 Ansatz & $x_B$ & $\Delta (x_B G)/x_B G$
& $\Delta (x_B G)/\Delta (x_B G)_{GLR}$ \\ \hline
MRSD0' & $10^{-2}$ & $-1.4\cdot 10^{-3}$ & $0.11$\\
       & $10^{-4}$ & $-8.9\cdot 10^{-3}$ & $0.29$\\
       & $10^{-5}$ & $-14.7\cdot 10^{-3}$ & $0.31$\\ \hline
MRSD-' & $10^{-2}$ & $-1.2\cdot 10^{-3}$   &$0.084$ \\
       & $10^{-4}$ & $-13.5\cdot 10^{-3}$  &$0.1$\\
       & $10^{-5}$ & $-40.5\cdot 10^{-3}$  &$0.09$\\ \hline
\end{tabular}
\vspace{0.4cm}

Table 2. Correction to the gluon distribution function
for two different {\it ans\"{a}tze}.
Here $Q^2 = 15$ ${\rm GeV}^2/c^2$, $\alpha_s \simeq 0.21$
and $R_N^2 = 5$ ${\rm GeV}^{-2}$.


\pagestyle{empty}

\centerline{\bf FIGURE CAPTIONS}

\parindent=0pt

{\bf Figure 1.}

Rescattering of Pomerons
in the t-channel.

\bigskip

{\bf Figure 2.}

Pictorial representation of the generalized evolution
equation.

\bigskip

{\bf Figure 3.}

``Fan'' diagram.

\bigskip

{\bf Figure 4.}

Production of three gluon shadows in a parton cascade.

\bigskip

{\bf Figure 5.}

Example of type of multigluon interactions that the generalized
evolution equation takes into account.

\newpage

\vfill\vbox{\includegraphics{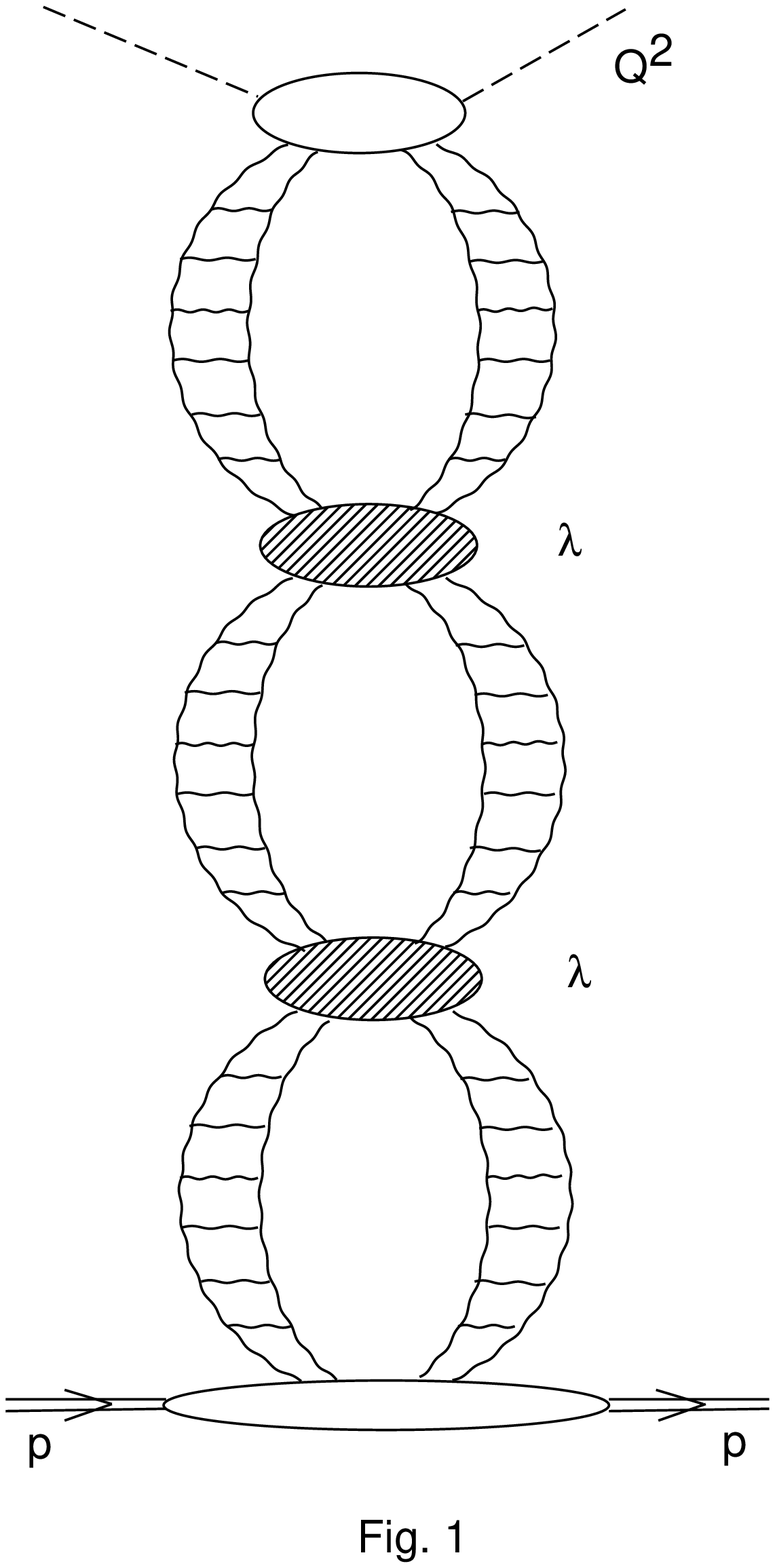}}

\newpage

\vbox{\includegraphics{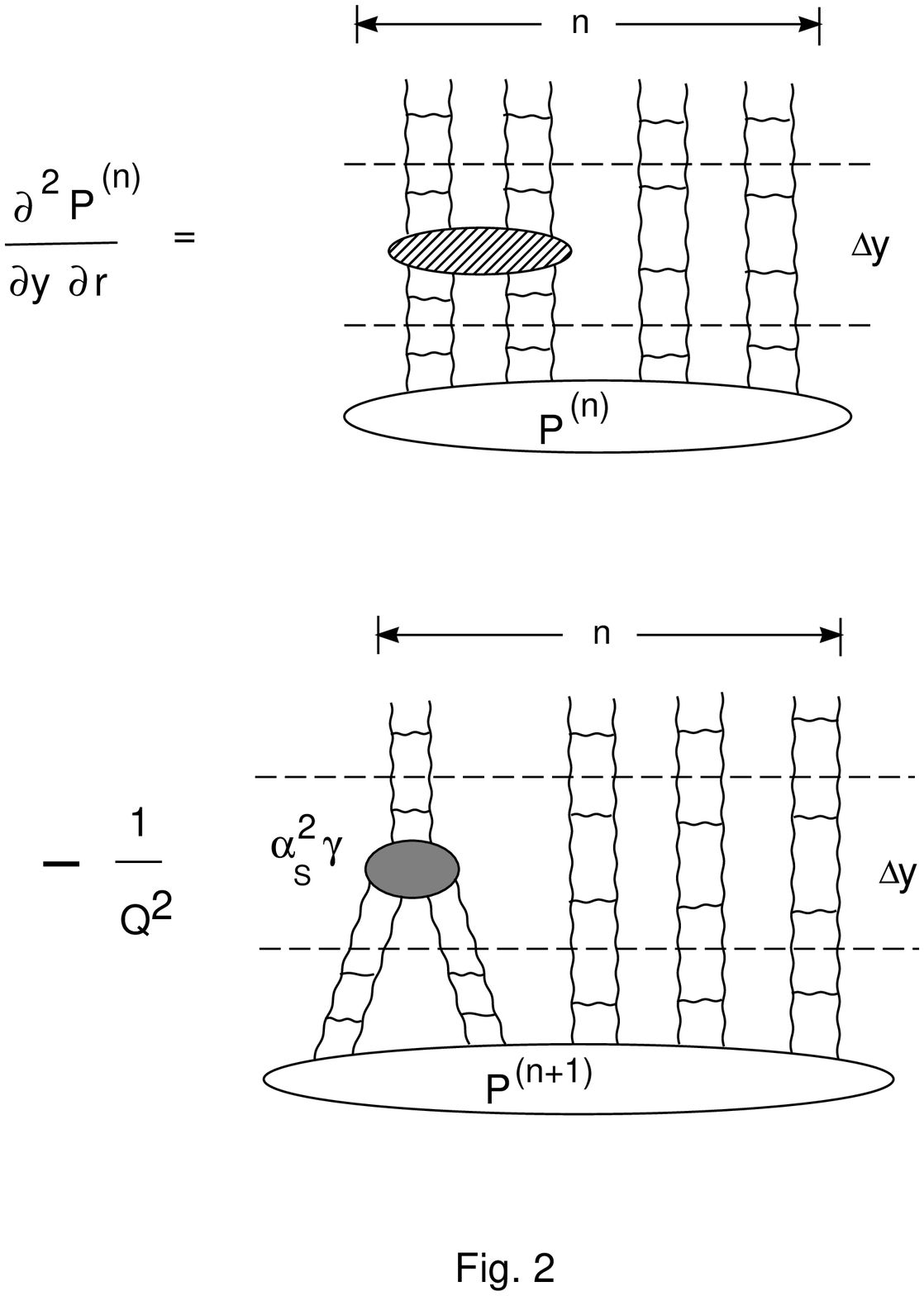}}

\newpage

\vbox{\includegraphics{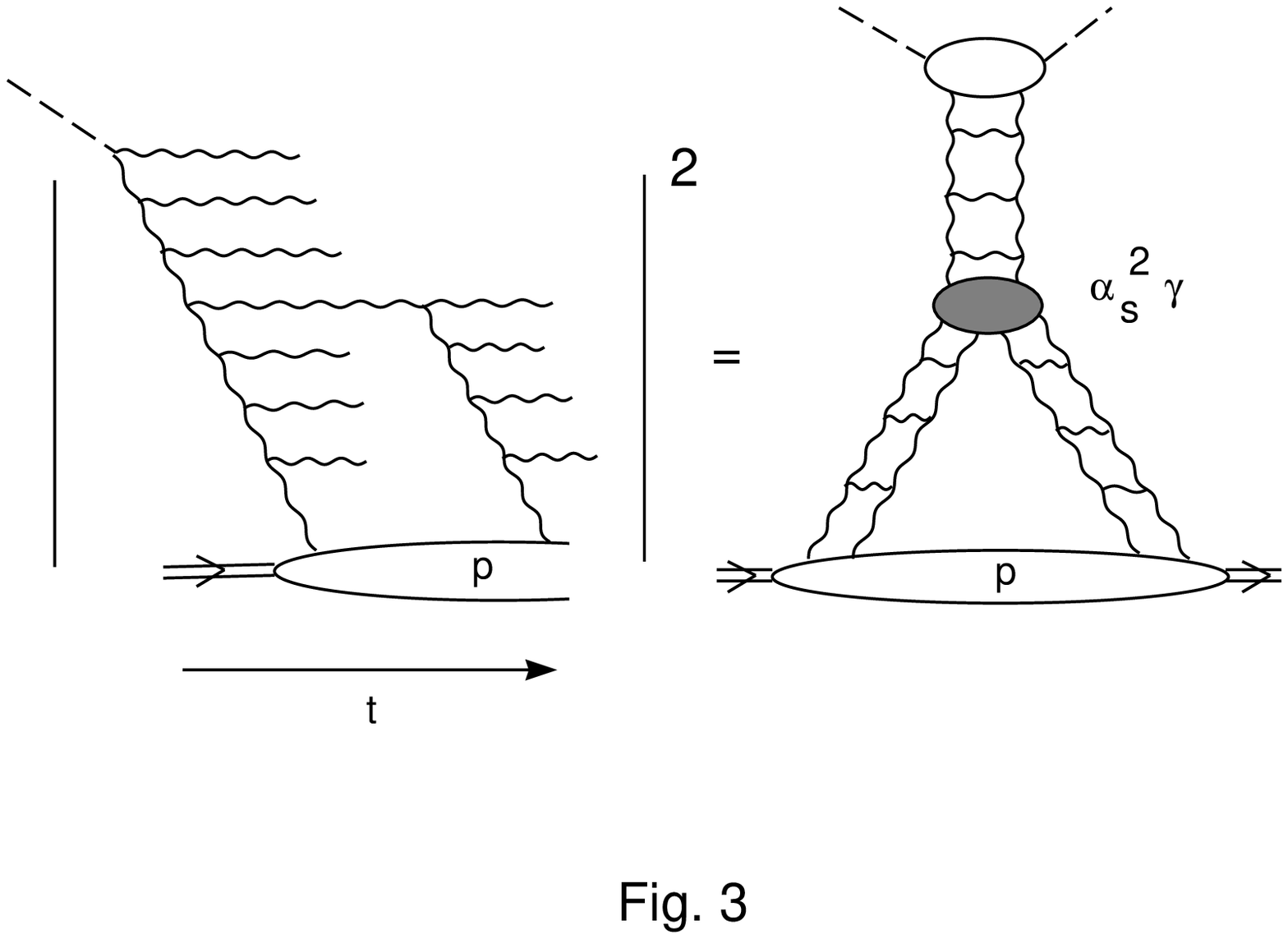}}
\newpage

\vbox{\includegraphics{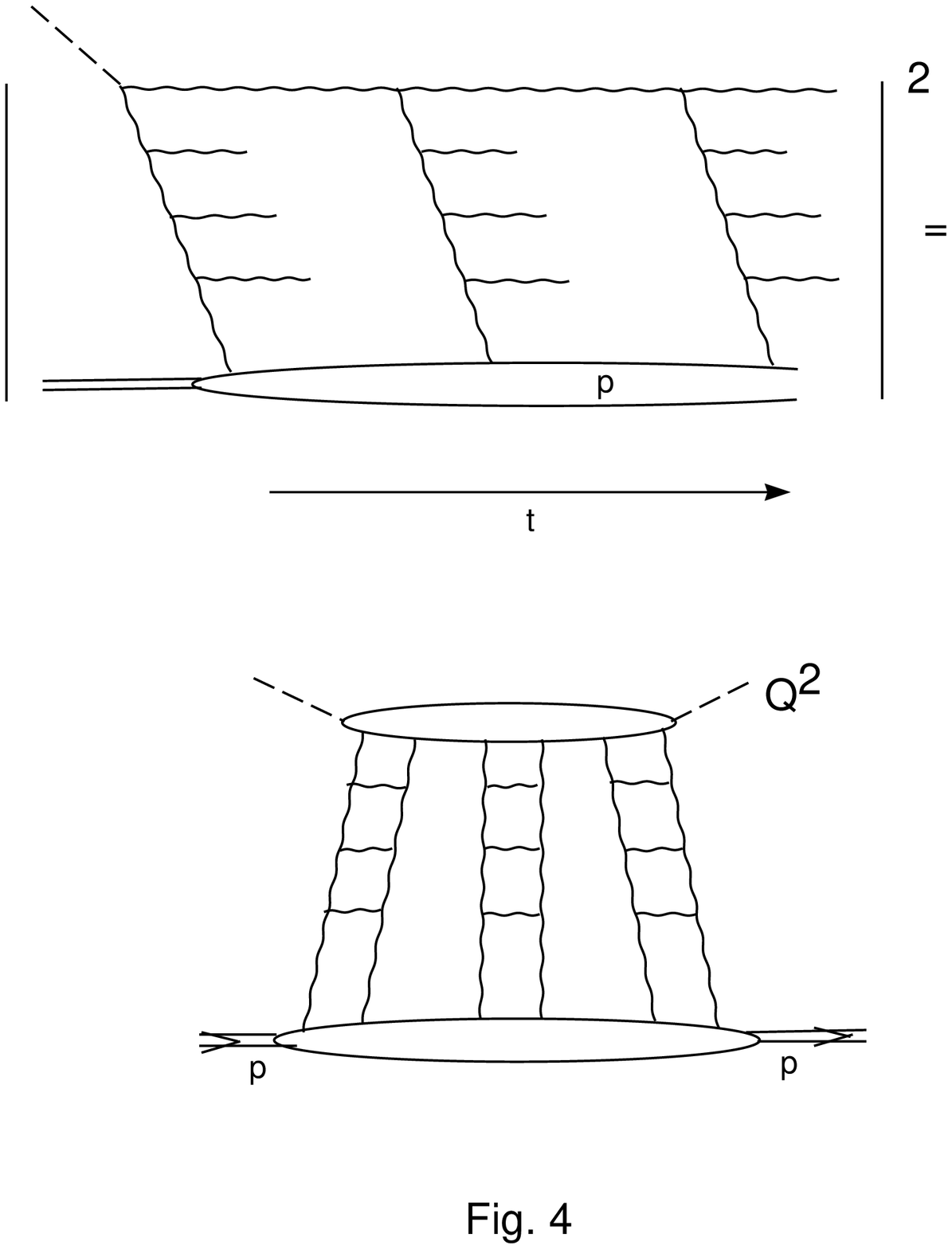}}
\newpage

\vbox{\includegraphics{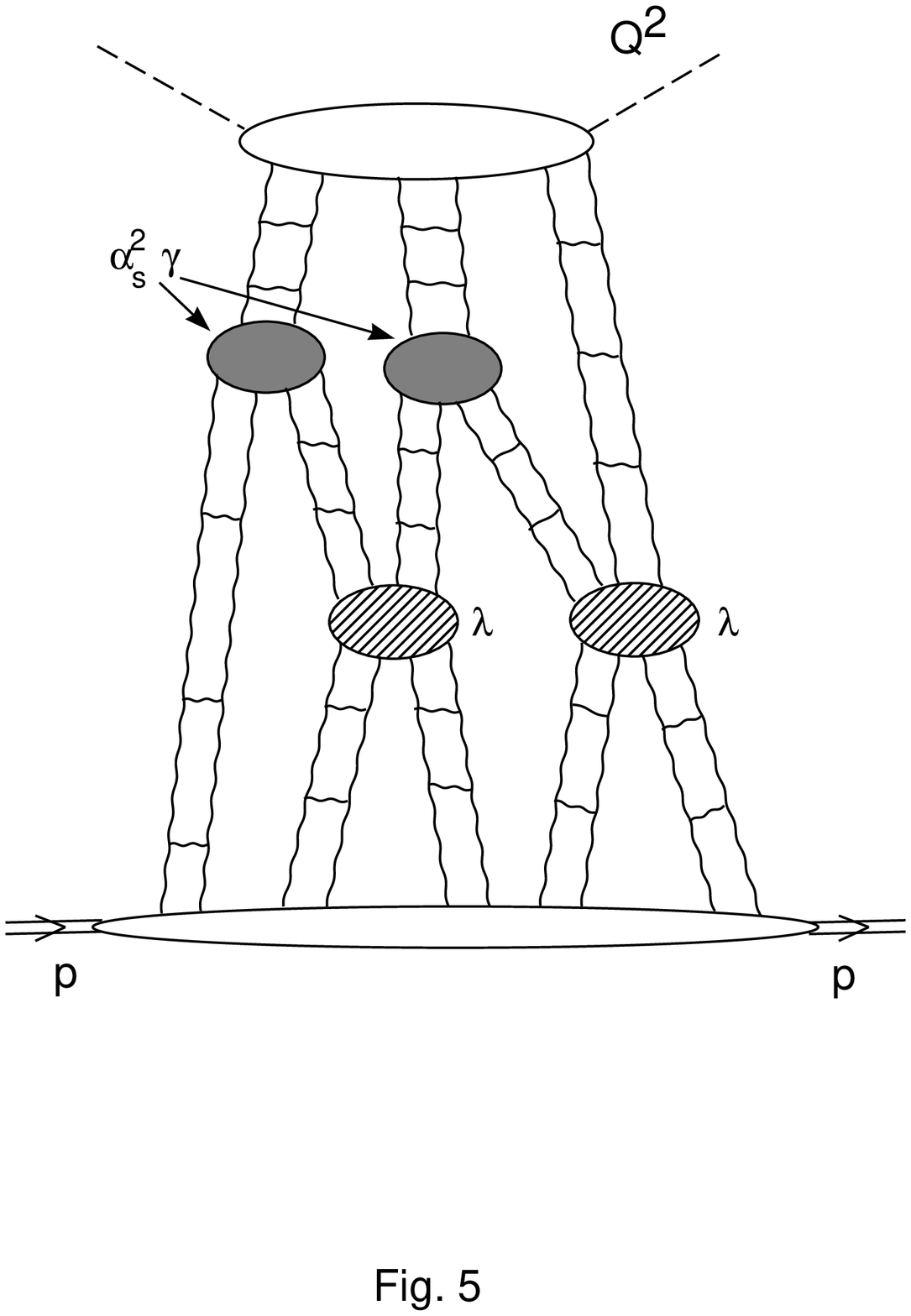}}

\end{document}